\documentclass[prl,amsmath,amsfonts,amssymb,twocolumn,showpacs]{revtex4}

\usepackage{graphicx,psfrag,color}
\usepackage{dcolumn}
\usepackage{bm}
\usepackage{mathbbol}

\begin{document}

\title{Thermal and Magnetic Quantum Discord in Heisenberg models}

\author{T. Werlang}
\affiliation{Departamento de Fisica, Universidade Federal de Sao
Carlos, Sao Carlos, SP 13565-905, Brazil}
\author{Gustavo Rigolin}
\affiliation{Departamento de Fisica, Universidade Federal de Sao
Carlos, Sao Carlos, SP 13565-905, Brazil}

\date{\today}

\begin{abstract}
We investigate how quantum correlations (quantum discord (QD)) of
a two-qubit one dimensional XYZ Heisenberg chain in thermal
equilibrium depend on the temperature T of the bath and also on an
external magnetic field B. We show that the behavior of thermal QD
differs in many unexpected ways from thermal entanglement. For
example, we show situations where QD increases with T when
entanglement decreases, cases where QD increases with T even in
regions with zero entanglement, and that QD signals a quantum
phase transition even at finite T. We also show that by properly
tuning B or the interaction between the qubits we get non-zero QD
for any T and we present a new effect not seen for entanglement,
the ``regrowth'' of thermal QD.
\end{abstract}


\maketitle

\textit{Introduction.} Since the seminal work of John S. Bell
\cite{Bel64}, who brought to the realm of experimental Physics the
ideas of Einstein, Podolsky, and Rosen \cite{Ein35} concerning the
non-local aspects of Quantum Mechanics, it became clear that the
constituents of some quantum composite systems possessed
correlations among themselves unachievable in the classical world.
Those states belong to the class of entangled states although not
all entangled states possess stronger-than-classical correlations
in the sense of violating Bell inequalities \cite{Wer89}.
However,
given a composite quantum state whose constituents are correlated,
how can one tell the origin of the correlations? In other words,
how can one divide the total correlation into a classical part and
a purely quantum one? This is particularly important for mixed
states since their quantum correlations are many times hidden by
their classical correlations. An answer to these questions is
given by the quantum discord (QD), a measure of the quantumness of
correlations introduced in Ref. \cite{Zur02}. QD is built on the
fact that two classical equivalent ways of defining the mutual
information turn out to be inequivalent in the quantum domain. In
addition to its conceptual role, some recent results
\cite{Dat08} suggest that QD and not entanglement may be
responsible for the efficiency of a mixed state based quantum
computer.

Due to its fundamental and practical significance we wish to
investigate the amount of QD in a concrete system such as a pair
of qubits (spin-1/2) within a solid at finite temperature, whose
interaction is given by the Heisenberg model (XYZ model in
general). Such Heisenberg models can describe fairly well the
magnetic properties of real solids \cite{Ham99} and is well
adapted to the study of the interplay of disorder and entanglement
as well as of entanglement and quantum phase transitions
\cite{Rig}. We characterize the dependence of QD on the
temperature and also on external magnetic fields applied to the
qubits. We also compare the behavior of QD against the
entanglement of formation (EoF) between the two qubits
\cite{Ved01,Kam02,Rig03}. We obtain several interesting and novel
results for the behavior of QD \textit{at non-zero T}, many of
them in contrast to the behavior of EoF. First, we show that QD
can \textit{increase} with temperature in the \textit{absence} of
external fields applied to the qubits. This is in sharp contrast
to the behavior of EoF since one can show \cite{Rig03} that this
effect never occurs without the presence of an external magnetic
field. We also show that there exist regions
where QD is different from zero while EoF is always zero, 
confirming a generic feature of QD \cite{Fer09}. 
In particular, we show that for the isotropic XXX model,
both the ferromagnetic and anti-ferromagnetic Hamiltonians possess
relatively high values of QD while EoF is absent for the
ferromagnetic model \cite{Ved01}. These properties of QD can have
important practical consequences for the realization of a quantum
computer at finite temperatures. Indeed, once it
is established that what is at stake for the correct functioning
of a quantum computer is the existence of a certain level of QD
(just having non-null QD might not be enough since almost all
states have QD$>0$ \cite{Fer09}), we have shown that Heisenberg
solids are better than previously thought as good candidates for
the construction of a quantum computer at $T>0$: for some
parameter sets they have high values of QD while having no EoF at
all.  We also show that by properly adjusting the coupling
constants \cite{Kam02} one can achieve any desired level of QD for
any T. Finally, we show that for finite T we observe the
sudden-change of QD \cite{Maz09} as we change the coupling
constant of the Hamiltonian and we show a new effect, namely the
``regrowth" of the thermal QD with increasing T.

\textit{The thermalized Heisenberg system.} The Hamiltonian of the
XYZ model with an external magnetic field acting on both qubits is
\begin{equation}
H = B(S_{z}^{1} + S_{z}^{2}) + J_xS_{x}^{1}S_{x}^{2}+
J_yS_{y}^{1}S_{y}^{2} + J_{z}S_{z}^{1}S_{z}^{2}, \label{xyz2}
\end{equation}
with $J_{x}, J_{y}$, and $J_{z}$ the coupling constants,
$S^j_{x,y,z} = \sigma^j_{x,y,z}/2$,  $\sigma_{x}^{j},
\sigma_{y}^{j}$, and $\sigma_{z}^{j}$ the usual Pauli matrices
acting on qubit $j$, and $B$ the external magnetic field. We have
assumed $\hbar = 1$.
The density matrix describing a system in equilibrium with a
thermal reservoir at temperature T (canonical ensemble) is $\rho =
\exp{\left( -H/kT  \right)}/Z$, where $Z = \mbox{Tr}\left\{
\exp{\left( -H/kT  \right)} \right\}$ is the partition function
and $k$ is Boltzmann's constant. Therefore, Eq.~(\ref{xyz2}) leads
to the following thermal state in the standard basis
%
\begin{equation}
\rho  =  \frac{1}{Z} \left(
\begin{array}{cccc}
 A_{11} & 0 & 0 & A_{12}\\
0 & B_{11}  & B_{12} & 0 \\
0 & B_{12} & B_{11} & 0 \\
A_{12} & 0 & 0 & A_{22} \\
\end{array}
\right). \label{rho}
\end{equation}
Here $A_{11}$  $=$ $\mathrm{e}^{-\alpha}$ $(\cosh(\beta)$ $-$ $4B$
$\sinh(\beta)/\eta)$, $A_{12}$ $=$  $-$ $\Delta$
$\mathrm{e}^{-\alpha}$ $\sinh(\beta)/\eta$, $A_{22}$ $=$
$\mathrm{e}^{-\alpha}$ $(\cosh(\beta)$ $+$ $4$ $B$
$\sinh(\beta)/\eta)$, $B_{11}$ $=$ $\mathrm{e}^{\alpha}$
$\cosh(\gamma)$, $B_{12}$ $=$ $-$ $\mathrm{e}^{\alpha}$
$\sinh(\gamma)$,
and $Z = 2\left( \exp{(-\alpha)}\cosh(\beta) +
\exp{(\alpha)}\cosh(\gamma) \right)$, where $\Delta = J_{x} -
J_{y}$, $\Sigma = J_{x} + J_{y}$, $\eta = \sqrt{\Delta^{2} +
16B^2}$, $\alpha = J_{z}/(4kT)$, $\beta = \eta/(4kT)$, and $\gamma
= \Sigma/(4kT)$.

\textit{Entanglement.} For a pair of qubits there exists an
analytical expression to quantify its amount of entanglement
called Entanglement of Formation (EoF) \cite{Ben96}. Given the
density matrix $\rho$ describing thermalized two qubits, EoF is
the average entanglement of the pure state decomposition of
$\rho$, minimized over all possible decompositions,
%
$ EoF(\rho) = \mbox{min}\sum_{k}p_{k}E(\phi_{k}), $
%
where $\sum_{k}p_{k} = 1$, $0 < p_{k} \leq 1$, and $\rho =
\sum_{k}p_{k}\left| \phi_{k}\right>\left< \phi_{k}\right|$.
$E(\phi_{k})$ is the entanglement of the pure state
$|\phi_{k}\rangle$ \cite{Ben96b}. For a pair of qubits Wootters
\cite{Woo98} has shown that EoF is a monotonically increasing
function of the concurrence $C$ (an entanglement monotone), $ EoF
= -f(C)\log_2f(C) - (1-f(C))\log_2(1-f(C))$, where $f(C)=(1 +
\sqrt{1 - C^2})/2$.
The concurrence is
simply\cite{Woo98}
%
$
C = \mbox{max} \{ 0, \lambda_{1} - \lambda_{2} - \lambda_{3} -
\lambda_{4} \},
$
%
where $\lambda_{1},  \lambda_{2}, \lambda_{3}$, and $\lambda_{4}$
are the square roots of the eigenvalues, in decreasing order, of
the matrix $R =\rho\tilde{\rho}$. Here $\tilde{\rho}$ is the time
reversed matrix 
%
$ \tilde{\rho} = \left(\sigma_{y}^1 \otimes \sigma_{y}^2 \right)
\rho^{*} \left(\sigma_{y}^1 \otimes \sigma_{y}^2 \right). $
%
The symbol $\rho^{*}$ means complex conjugation of the matrix
$\rho$ in the standard basis $\left\{ \left| 00 \right>, \left| 01
\right>, \left| 10 \right>, \left| 11 \right> \right\}$. For a
density matrix in the X-form above
$C=2\rm{max}\{0,\Lambda_1,\Lambda_2\}/Z$, with
$\Lambda_1=|B_{12}|-\sqrt{A_{11}A_{22}}$ and
$\Lambda_2=|A_{12}|-B_{11}$.

\textit{Quantum Discord.} In classical information theory (CIT)
the total correlation between two systems (two sets of random
variables) $\mathcal{A}$ and $\mathcal{B}$ described by a joint
distribution probability $p(\mathcal{A},\mathcal{B})$ is given by
the mutual information (MI),
\begin{equation}
\mathcal{I}(\mathcal{A},\mathcal{B})=H(\mathcal{A})+
H(\mathcal{B})-H(\mathcal{A},\mathcal{B}),  \label{Tcorrelation}
\end{equation}
with the Shannon entropy $H(\cdot)=-\sum_{j}p_{j}\log_2 p_{j}$.
Here $p_j$ represents the probability of an event $j$ associated
to systems $\mathcal{A}$, $\mathcal{B}$, or to the joint system
$\mathcal{A} \mathcal{B}$. Using Bayes's rule we may write MI as
\begin{equation}
\mathcal{I}(\mathcal{A},\mathcal{B})=
H(\mathcal{A})-H(\mathcal{A}|\mathcal{B}),  \label{Tcorrelation2}
\end{equation}
where $H(\mathcal{A}|\mathcal{B})$ is the classical conditional
entropy. In CIT these two expressions are equivalent but in the
quantum domain this is no longer true \cite{Zur02}. The first
quantum extension of MI, denoted by
${\mathcal{I}}\left(\rho\right)$, is obtained directly replacing
the Shannon entropy in (\ref{Tcorrelation}) with the von Neumann
entropy,
$S\left(\rho\right)=-\mathrm{Tr}\left(\rho\log_2\rho\right)$, with
$\rho$, a density matrix, replacing probability distributions. To
obtain a quantum version of (\ref{Tcorrelation2}) it is necessary
to generalize the classical conditional entropy. This is done
recognizing $H(\mathcal{A}|\mathcal{B})$ as a measure of our
ignorance about system $\mathcal{A}$ after we make a set of
measurements on $\mathcal{B}$. When $\mathcal{B}$ is a quantum
system the choice of measurements determines the amount of
information we can extract from it. We restrict ourselves to von
Neumann measurements on $\mathcal{B}$ described by a complete set
of orthogonal projectors, $\left\{\Pi_j\right\}$, corresponding to
outcomes $j$. After a measurement, the quantum state $\rho$
changes to $\rho_{j}=\left[\left(\mathbb{I}\otimes\Pi_{j}\right)
\rho\left(\mathbb{I}\otimes\Pi_{j}\right)\right]/p_j$, with
$\mathbb{I}$ the identity operator for system $\mathcal{A}$ and
$p_j=\mathrm{Tr}[(\mathbb{I}\otimes\Pi_{j})
\rho(\mathbb{I}\otimes\Pi_{j})]$. Thus, one defines the quantum
analog of the conditional entropy as $S\left(\rho\mid\left\{
\Pi_{j}\right\}\right)=\sum_{j}p_{j}S\left(\rho_{j}\right)$ and,
consequently, the second quantum extension of the classical MI as
\cite{Zur02}
$\mathcal{J}\left(\rho\mid\left\{\Pi_{j}\right\}\right)=
S\left(\rho^{\mathcal{A}}\right)-S\left(\rho\mid\left\{\Pi_{j}\right\}
\right)$. The value of
$\mathcal{J}\left(\rho\mid\left\{\Pi_{j}\right\}\right)$ depends
on the choice of $\left\{\Pi_j\right\}$. Henderson and Vedral
\cite{vedral} have shown that the maximum of
$\mathcal{J}\left(\rho\mid\left\{\Pi_{j}\right\}\right)$ with
respect to $\left\{\Pi_j\right\}$ can be interpreted as a measure
of classical correlations. Therefore, the difference between the
total correlations ${\mathcal{I}}\left(\rho\right)$ and the
classical correlations $\mathcal{Q}\left(\rho\right)=\sup_{\left\{
\Pi_{j}\right\}}\mathcal{J}\left(\rho\mid\left\{ \Pi_{j}\right\}
\right)$ is defined as
\begin{eqnarray}\label{discord}
D\left(\rho\right)={\mathcal{I}}\left(\rho\right)-\mathcal{Q}\left(\rho\right),
\end{eqnarray}
giving, finally, a measure of quantum correlations \cite{Zur02}
called quantum discord (QD). For pure states QD reduces to entropy
of entanglement \cite{Ben96b}, highlighting that in this case all
correlations come from entanglement. However, it is possible to
find separable (not-entangled) mixed states with nonzero QD
\cite{Zur02}, meaning that entanglement does not cause all
nonclassical correlations contained in a composite quantum system. 
Also, QD can be operationally seen as the 
difference of work that can be extracted from a heat bath using a bipartite
system acting either globally or only locally \cite{Zur03}.
For more insights into QD the reader is referred to \cite{Hor05}.
In this work, when $B=0$ the density operator (\ref{rho}) is such
that QD is given by an analytical expression obtained in
\cite{luo}. When $B\neq0$ we computed QD numerically
\cite{werlang}.

\begin{figure}[!ht]
\includegraphics[width=6cm]{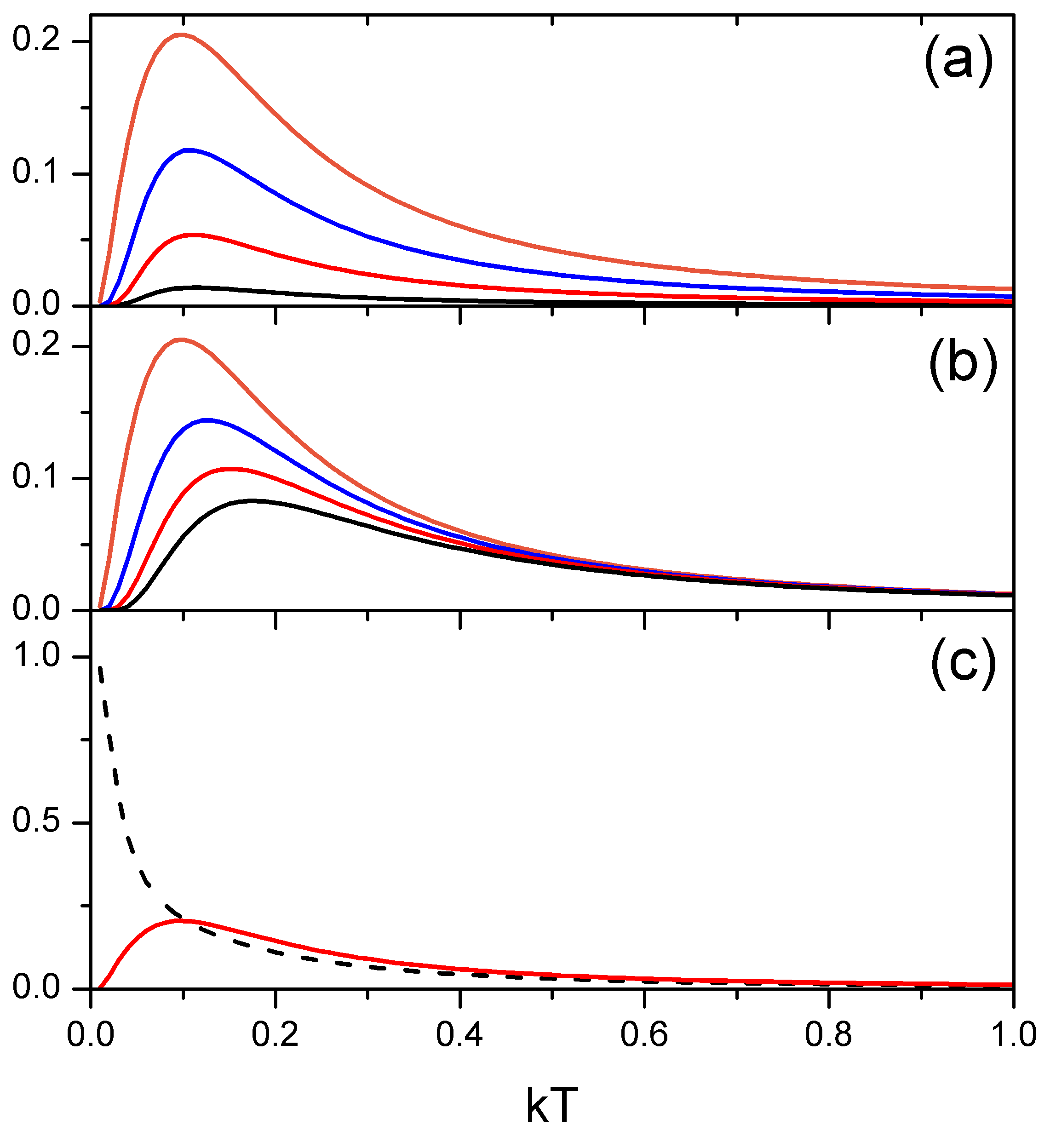}
\caption{\label{Fig1} (Color online) (a) and (b) QD as a function
of the absolute temperature $kT$ for the $XXZ$ model with $B=0$.
(a) Here $J_z=-0.5$ and from bottom to top $J$ $=$ $0.1, 0.2, 0.3,
0.4$; (b) Now we fix $J=0.4$ and from bottom to top
$J_z=-0.8,-0.7,-0.6,-0.5$. (c) QD (solid/red line) and classical
correlations, $\mathcal{Q}\left(\rho\right)$, (dashed/black line),
as functions of $kT$ for $J=0.4$ and $J_z=-0.5$. Here and in the
following graphics the quantities plotted are dimensionless.}
\end{figure}

\textit{Results.} Let us start presenting the important result
that QD increases with temperature without an external field
acting on the qubits ($B=0$). This effect can be clearly seen when
we deal with the XXZ model ($J_x=J_y=J$ and $J_z\neq 0$). Looking
at Fig. \ref{Fig1}, panels $a$ and $b$, we see that this effect
happens for several configurations of coupling constants being,
thus, dense around this region. We should note that such behavior
can be found for other models than the XXZ as well.
As we mentioned before, in the absence of
external fields \cite{Rig03} such increase with T only occurs for
QD and not for EoF. Note also that QD starts at zero and then
increases with T. In particular, for the set of coupling constants
shown in Fig. \ref{Fig1}, the EoF is always zero \cite{Rig03}.
Furthermore, one can show that the classical correlation decreases
when QD increases (See Fig. \ref{Fig1}c).  We have, therefore, a
genuine increase of quantum correlations with decreasing classical
correlation. This non trivial and unexpected effect indicates the
robustness of quantum correlations against classical correlations
as we increase the temperature.

\begin{figure}[!ht]
\includegraphics[width=6cm]{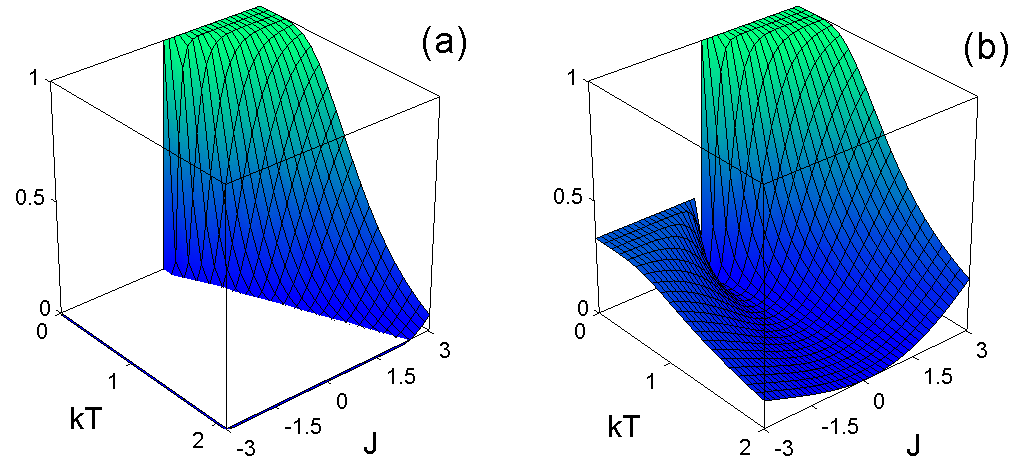}
\caption{\label{Fig1b} (Color online) EoF (a) and QD (b) as a
function of the absolute temperature $kT$ and the coupling $J$.
Both plots for the XXX model with $B=0$.}
\end{figure}

We now move to the $XXX$ model ($J_x=J_y=J_z=J$) with no field.
Looking at Fig. \ref{Fig1b} we see that although EoF is zero in
the ferromagnetic region ($J<0$) we have non-zero values for QD.
Also, both in the ferro and antiferromagnetic regions QD increases
if we increase the absolute value of $J$. Actually, for any finite
T we can find a $J$ big enough such that  QD $\neq 0$. This point
is justified observing that for any finite value of $T$, the
density operator (\ref{rho}) when $J\rightarrow\infty$ is given by
the Bell state $\rho=\left|\psi\right\rangle\left\langle
\psi\right|$ with
$\left|\psi\right\rangle=\frac{1}{\sqrt{2}}\left(\left|01\right\rangle
-\left|10\right\rangle\right)$. Besides, the density operator in
the limit $J\rightarrow-\infty$ for a finite $T$ is a mixed state
$\rho=\frac{1}{3}\left(\left|00\right\rangle\left\langle
00\right|+\left|11\right\rangle\left\langle
11\right|+\left|\phi\right\rangle\left\langle \phi\right|\right)$
with
$\left|\phi\right\rangle=\frac{1}{\sqrt{2}}\left(\left|01\right\rangle
+\left|10\right\rangle\right)$. The value of QD in this case is
$1/3$ and, as expected, the EoF is zero. We should also mention
that QD behaves qualitatively differently for $T\geq 0$ whether
$J<0$ or $J>0$, achieving the value zero at the critical point
$J=0$. This suggests that QD can signal the
quantum phase transition (QPT) taking place at $T=0$ (when we tune
$J$ driving the system from its anti to its ferromagnetic phase)
even at finite $T$. Note that EoF does not signal this QPT at
finite T: EoF goes to zero \textit{before} the
critical point for $T>0$ while $QD=0$ exactly at the critical
point, and only there. However, just as was done for $T=0$
\cite{qpt}, a detailed study of the behavior of QD and QPT for
$T>0$ is needed.

\begin{figure}[!ht]
\includegraphics[width=6cm]{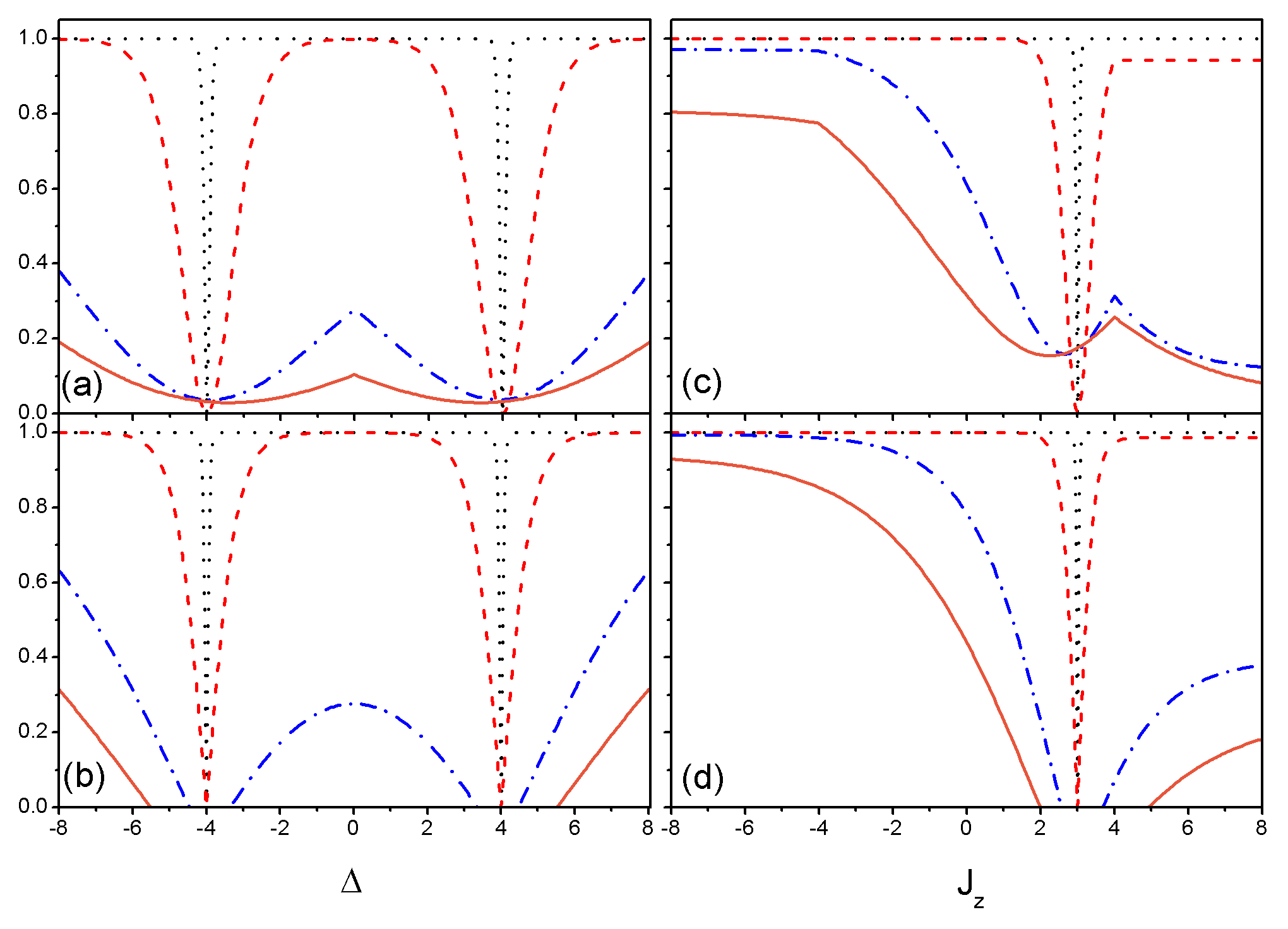}
\caption{\label{Fig2} (Color online) QD (a) and EoF (b) as a
function of $\Delta=J_x - J_y$ for $J_x+J_y=2$, $J_z=1$, and
various values of $kT$. QD (c) and EoF (d) as a function of $J_z$
for $J_x+J_y=1$, $\Delta=7$, and various values of $kT$. For the
dotted/black line $kT=0.01$, dashed/red line $kT=0.1$,
dash-dotted/blue line $kT=0.6$, and solid/orange line $kT=1$. }
\end{figure}

Focusing now on the XYZ model we investigate how QD behaves while
we change the coupling constants for a fixed temperature T. At the
left panels of Fig. \ref{Fig2} we plot how QD and EoF depend on
the anisotropy of the $J_x$ and $J_y$ coupling constants for $J_z$
fixed. We note that after a certain T the derivatives of QD with
respect to $\Delta = J_x - J_y$ are undefined at $\Delta = 0$. This
does not happen to EoF. The discontinuity of the derivative shown
here is similar to what was observed to the dynamical decay rate
of QD under decoherence in Ref. \cite{Maz09}, where the authors
coined the name ``sudden change'' for this behavior \cite{Guo09}.
At the right panels of Fig. \ref{Fig2} we see the sudden change of
QD keeping all parameters fixed but $J_z$. As before, we do not
see sudden change for the EoF. The sudden change
of QD with the coupling constants here and the one with time 
in \cite{Maz09}
suggest that this behavior may be related to a technicality
of QD rather than to a change in a physical property of the system. 
However, no proof of this point is given and further investigation
is needed.

\begin{figure}[!ht]
\includegraphics[width=6cm]{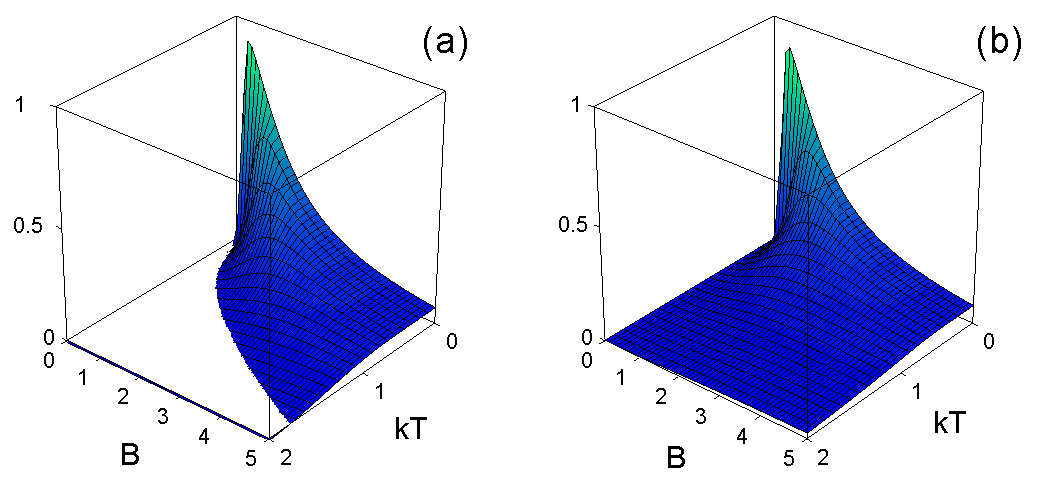}
\caption{\label{Fig5} (Color online) EoF (a) and QD (b) as a
function of the magnetic field B and the absolute temperature kT
for the Ising model when $J=1$. }
\end{figure}

When the magnetic field $B$ is not zero we first analyze what
happens to the Ising model ($J_x=J$ and $J_y=J_z=0$). Looking at
Fig. \ref{Fig5} we see that for the regions where EoF is zero QD
is negligible. However, as well as with EoF \cite{Kam02}, QD
initially increases as we increase the value of $B$, going to zero
with increasing field. This is true since the density operator
(\ref{rho}) is $\left|11\right\rangle\left\langle 11\right|$
(separable state) when $B\rightarrow\infty$. On the other hand, as
we will shortly see, this is not a general result. The behavior
of QD and EoF can be quite different from each other if we work
with another model.

\begin{figure}[!ht]
\includegraphics[width=6cm]{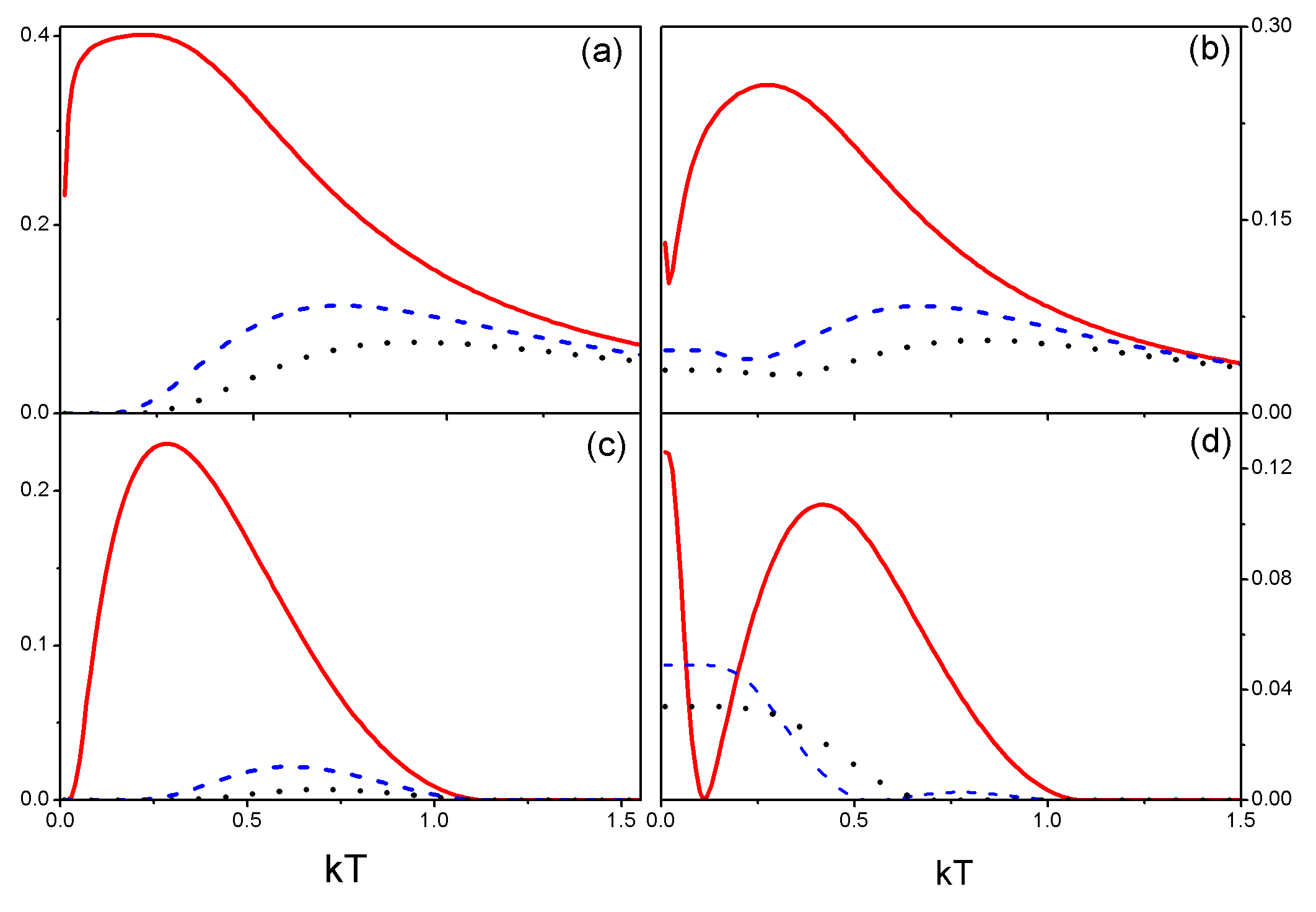}
\caption{\label{Fig3} (Color online) QD, panels (a) and (b), and
EoF, panels (c) and (d), as a function of the absolute temperature
$kT$ for the XY model with transverse magnetic field $B$. For (a)
and (c) $J_x=J_y=1$ and for (b) and (d) $J_x=1.3$ and $J_y=0.7$.
The values for $B$ are $1.1$ (red/solid line), $2.0$ (blue/dashed
line), and $2.5$ (black/dotted line).}
\end{figure}

For the XY model ($J_x$, $J_y\neq 0$ and $J_z=0$) with a
transverse magnetic field ($B\neq 0$), for example, QD behaves in
quite different (and interesting) ways from what we have seen for
the Ising model. Also, for the XY model we see a new effect, the
\textit{regrowth} of QD. Contrary to the behavior of EoF, where we
see its sudden death and then a revival \cite{Kam02}, for QD there
is no sudden death. See Fig. \ref{Fig3}. QD decreases with T,
retaining appreciable values and then after a critical temperature
$T_c$ it starts increasing again. This is what we call regrowth.
Note that in the behavior of EoF we never see a regrowth. Indeed,
EoF increases after decreasing with T only after reaching the
value zero (sudden death).

Moreover, looking at (a) and (c) of Fig. \ref{Fig3} we see
that QD becomes non-null before the appearance of EoF. Also, QD
continues to be non-null after EoF disappears. If we now analyze
(b) and (d) of Fig. \ref{Fig3} we notice that we have for
all three curves regimes in which QD increases while EoF
decreases. This is a quite remarkable behavior since the decrease
of a certain quantum aspect (entanglement) is simultaneous to the
increase of another quantum aspect (quantum correlations),
illustrating clearly the distinctive aspects of these two
concepts.

\textit{Conclusions.} We examined the behavior of the quantum
discord (QD) for a pair of qubits described by the Heisenberg
model in thermal equilibrium with a reservoir at temperature T. By
changing the temperature and also by applying an external magnetic
field we observed several remarkable effects for QD, many of them
in sharp contrast to the behavior observed for the entanglement of
formation (EoF) between the two qubits. We found that for the XXX
model QD signals a quantum phase transition for finite T while EoF
does not. Also, we observed regimes where QD increases while EoF
decreases with T. Moreover, and surprisingly, we showed that for
the XXZ model there exist regions in the parameter space in which
EoF is zero while the classical correlation decreases and only QD
increases with T.
Finally, we also observed a new effect for QD which we called
\textit{regrowth}: QD decreases with T and starts increasing again
after reaching a minimum different from zero.

\begin{acknowledgments}
T.W. thanks CNPq for partially funding this research.
\end{acknowledgments}

\end{document}